\documentclass[twocolumn,prl,aps
,showpacs,preprintnumbers,amsmath
,amssymb]{revtex4}

\usepackage{graphicx}% Include figure files
\usepackage{dcolumn}% Align table columns on decimal point
\usepackage{bm}% bold math
\usepackage{subfigure} 
\usepackage{array}
\usepackage{slashbox}
\usepackage{psfrag}
\usepackage{epstopdf}

\begin{document}

\title{Vortices near the Mott phase of a trapped Bose-Einstein condensate}

\author{Daniel S. Goldbaum}
\email{dsg28@cornell.edu}
\author{Erich J. Mueller}\affiliation{
Laboratory of Atomic and Solid State Physics, Cornell University\\
Ithaca, NY 14853
}

\date{\today}


\newcommand{\adi}{\hat{a}_{i}^{\dagger} }
\newcommand{\ai}{\hat{a}_{i} }
\newcommand{\adj}{\hat{a}_{j}^{\dagger} }
\newcommand{\aj}{\hat{a}_{j} }
\newcommand{\numi}{\hat{n}_{i} }

\begin{abstract}{
We present a theoretical study of vortices within a harmonically trapped Bose-Einstein condensate in a rotating optical lattice.  We find that proximity to the Mott insulating state dramatically effects the vortex structures. To illustrate we give examples in which the vortices: (i) all sit at a fixed distance from the center of the trap, forming a ring, or (ii)  coalesce at the center of the trap, forming a
giant vortex. We model time-of-flight expansion to demonstrate
the experimental observability of our predictions.
}
\end{abstract}

\maketitle

Atomic clouds in a
 rotating optical lattice are at the intellectual intersection of several major paradigms of condensed matter physics. These rotating clouds may display a superfluid-insulator quantum phase transition~\citep{Greiner:2002lr}, vortex pinning~\citep{reijnders:060401}, frustration~\cite{Polini_LPhys_603_2004}, Josephson junction physics~\citep{polini:010401}, and even analogs of the fractional quantum Hall effect~\citep{0953-8984-20-12-123202}.  Here we explore the theory of vortices in such systems, showing how proximity to the Mott insulator phase impacts the vortex configurations.
  
 Considering a uniform gas of atoms of mass $m$ 
 in an optical lattice rotating with frequency $\Omega$, there are three
macroscopic length scales in the problem: the lattice spacing $d$, the  magnetic length $\ell=\sqrt{\hbar/m\Omega}$, and the particle spacing $n^{-1/3}$, where $\hbar=h/2\pi$ is Planck's constant.  Even without interactions, the commensurability of these lengths leads to nontrivial physics -- the single particle spectrum, the Hofstadter butterfly, is fractal~\citep{PhysRevB.14.2239}.  For interacting bosons, this fractal spectrum leads to a modulation of the boundary between  superfluid and Mott insulating phases~\citep{goldbaum:033629,umucalilar:055601}.  Further, the vortices in a superfluid on a rotating lattice develop extra structure:  their cores may fill with the Mott state~\citep{wu:043609},  changing which vortex arrangements minimize the energy~\citep{goldbaum:033629}.  

We consider a harmonically trapped superfluid gas on a rotating optical lattice in the single-band tight binding-limit close to the Mott state. We choose to study a two-dimensional cloud, as it provides the simplest setting for investigating vortex physics, and is an experimentally relevant geometry~\citep{spielman:120402}. 
The proximity to the Mott state results in a nontrivial spatial dependance of the superfluid order parameter, and drives a rearrangement of vortices. 

A similar geometry was realized in a recent experiment~\citep{tung:240402}, with the caveat that their  shallow optical lattice had such a large lattice spacing that they were not able to reach the tight binding limit.  In principle their technique can be refined to explore the physics that we describe here. The tight binding limit may also be reached through
 quantum optics techniques which introduce phases on the hopping matrix elements for atoms in a non-rotating lattice~\citep{JakschNJP2003}. 
 
In the rotating frame our system is described by 
the rotating Bose-Hubbard hamiltonian~\citep{wu:043609,PhysRevLett.81.3108}:
\begin{equation}
\hat{H}
=-\sum_{\langle i,j \rangle} \left(t_{ij} \adi  \aj  + h.c. \right) 
+  \sum_i \left(\frac{U}{2} \numi \left( \numi - 1 \right) -\mu_i \numi\right)
\label{B}
\end{equation}
where $t_{ij}=t\exp\left[ i \int_{\vec{r}_j}^{\vec{r}_{i}} d\vec{r} \cdot \vec{A}(\vec{r}) \right]$ is the hopping matrix element from site $j$ to site $i$.  The rotation vector potential, which gives rise to the Coriolis effect, is $\vec{A}(\vec{r}) = \left( m/\hbar\right) \left(\vec{\Omega} \times \vec{r} \right)= \pi \nu \left( x \hat{y} -y \hat{x} \right)$, where $\nu$ is the number of circulation quanta per optical-lattice site.  The local chemical potential
 $\mu_i=\mu_0-m \left( \omega^2-\Omega^2 \right) r_i^2 /2$ includes the centripetal potential.  In these expressions,
 $\mu_0$ is the central chemical potential, $\omega$ is the trapping frequency,
 ${\vec r}_i$ is the position of site $i$,  
 $\adi$ $\left( \ai \right)$ is a bosonic creation (annihilation) operator,
$\numi = \adi \ai $ is the particle number operator for site $i$,  and $U$ is the particle-particle interaction strength.      
The connection between these parameters and experiment are given by Jaksch et al. \citep{PhysRevLett.81.3108}.
Here, and in the rest of the paper, we use units where the lattice spacing is unity.
    
Both the superfluid and Mott insulator can be approximated by 
a spatially inhomogeneous Gutzwiller product \emph{ansatz}~\citep{PhysRevLett.81.3108}, $\lvert \Psi_{GW} \rangle = \prod_{i=1}^{M} \left( \sum_n f_n^i \lvert n \rangle_i \right)$, 
where $i$ is the site index, $M$ is the total number of sites, $\lvert n \rangle_i$ is the n-particle occupation-number state at site $i$, and $f_n^i$ is the corresponding complex amplitude, with $\sum_n |f_n^i|^2=1$.
Despite the limitations of being a mean-field theory, the Gutzwiller approach compares well with exact methods, and strong coupling expansions~\citep{PhysRevB.40.546}. It has also been used extensively to understand experimental results~\citep{Greiner:2002lr,folling:060403}, 
and is well suited for studying the vortex physics that we consider here.

Using  $\lvert \Psi_{GW} \rangle$ as a variational {\em ansatz}, we minimize the energy with respect to the $\{f_n^i \}$.  We then extract the density $\rho_i=\sum_n n |f_n^i|^2$ and the condensate order parameter $\alpha_i=\langle \hat{a}_i \rangle= \sum_n \sqrt{n} \left(f_{n-1}^{i}\right)^* f_n^i$ at each site. The condensate density $\rho_i^c=\lvert \langle \hat{a}_i \rangle \rvert^2$ is equal to the superfluid density in this model, and is generally not equal to the density.

We use an iterative algorithm to  determine the $\{f_n^i\}$ which minimize the energy.  We use a square region with $L$ sites per side and hard boundary conditions. We find that we must take 
$L$ much larger than the effective cloud diameter so that our solutions do not depend on those boundaries.  Typically we use $40\leq L \leq 90$. 
For the simulation described in figure 1 we impose four-fold rotational symmetry, but from unconstrained simulations on smaller clouds we find that this constraint does not significantly change the phenomena. %For the simulations described here we impose four-fold rotational symmetry, but from unconstrained simulations on smaller clouds we find that this constraint does not significantly change the phenomena. 
Similar calculations were performed by Scarola and Das Sarma~\citep{scarola:210403} to analyze the case where the single-particle Mott state is surrounded by a rotating superfluid ring.

Since this mean-field theory is highly nonlinear we find that the iterative algorithm often converges to different solutions depending on the initial state we use. For the results shown here we first iterate to self-consistency in a parameter region where the solution is unique, then slowly change parameter values, using the result from the previous parameters as a seed.  One should see analogous results in an experiment where one adiabatically changes parameters.  As in such experiments  \citep{PhysRevLett.86.4443} we observe hysteresis.

We have performed a thorough investigation of a wide range of parameters and, as one would expect, we find that
a basic understanding of the trapped gas can be extracted from the phase diagram of the homogeneous system, where $\Omega=\omega=0$ [Figs. 1(a) and 2(a)]. For a sufficiently gentle trap, the gas looks locally homogeneous, and its density at any point $r$ can be approximated by that of a uniform system with chemical potential $\mu(r)=\mu_0-V(r)$.  
%We go beyond this local density approximation (LDA) in our calculations, however it is useful for gaining qualitative insight into where interesting physics can be found. 
As a general rule, nontrivial vortex structures appear when the LDA superfluid density deviates significantly from a typical Thomas-Fermi profile.    The vortices tend to move to regions where there is a local suppression of the superfluid density.

\begin{figure}
\includegraphics[width=1.\columnwidth]{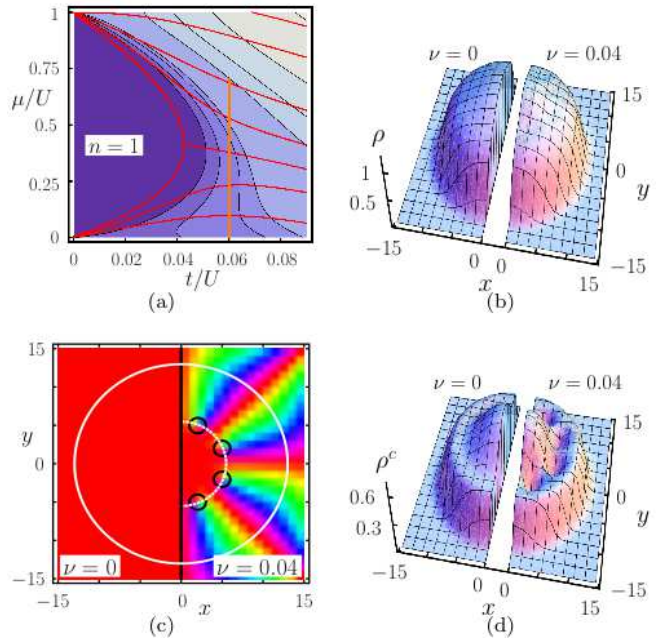}
\caption{\textbf{Ring vortex configuration} (color online, one-column). Comparison between non-rotating $\left( \nu=0 \right)$ and rotating $\left( \nu=0.04 \right)$ states of a system characterized by $\left(t/U=0.06,\,  \mu_0/U=0.7\right)$. (a) Mean-field phase plot of the uniform Bose-Hubbard model. Contours of fixed $\rho$ and $\rho^c$, are indicated by red and black curves. The superfluid density vanishes in the single-particle Mott region labeled ``$n=1$'', and increases with lightening shades of purple. The vertical orange line represents the LDA parameter-space trajectory for the current system.  (b) [(d)] Comparison of density [condensate density].
(c) Comparison of order parameter complex phase field. The complex phase is represented by ``Hue''. Solid and dotted white lines are a guide to the eye. Black circles enclose singly-quantized vortices.
As seen in (c) and (d), vortices form in a circular pattern on the condensate density plateau; the density (b) changes only slightly due to rotation.    }
\label{RingFigurePrep}
\end{figure}

\begin{figure}
\includegraphics[width=1.\columnwidth]{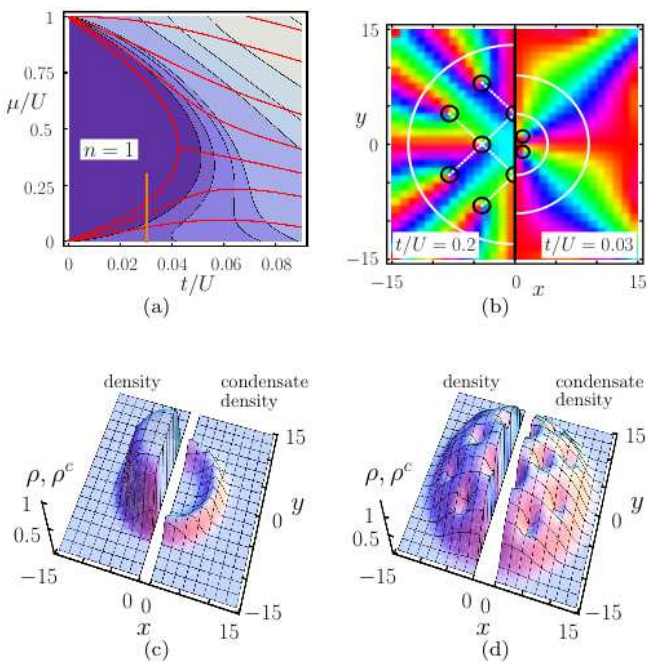}
\caption{\textbf{Giant vortex} (color online, one-column). Comparison of a vortex lattice far from the Mott regime $\left( t/U=0.2 \right)$ and a giant vortex system where the Mott phase occupies the center of the cloud $\left(t/U=0.03 \right)$. (a) Mean-field phase plot for the uniform Bose-Hubbard model with vertical orange parameter-space trajectory representing a system with $\left(t/U=0.03, \, \mu_0/U=0.3\right)$. (b) Comparison of order parameter complex phase fields. (c) [(d)] Comparison of density and condensate density where $t/U=0.03$ $\left[ t/U=0.2 \right]$.}
\label{GiantFigurePrep}
\end{figure}

We illustrate this principle with two examples:  in Fig.~\ref{RingFigurePrep} we study the case where the superfluid density has a ring-shaped plateau, and in Fig.~\ref{GiantFigurePrep} we consider the case where a Mott region sits in the center of the cloud.
 
%For the parameters used in Fig.~\ref{RingFigurePrep} we show the
%display the 
%density and complex phase profiles corresponding to a system characterized by the parameters $\left(t/U=0.06, \mu_0/U=0.7, R_{TF}=10%, L=50 
%\right)$, where $R_{TF}=\sqrt{\frac{2 \mu_0}{m\left( \omega^2-\Omega^2 \right)}}$ is the Thomas-Fermi radius for the rotating system. The cloud has a diameter of roughly 26 sites.  In the non-rotating case we see that the  superfluid density profile Fig. 1(d -- left) has a plateau structure (as one would predict from the LDA). No such plateau is seen in the total density Fig. 1(b -- left). The phase of the superfluid order parameter is uniform in the absence of rotation Fig. 1(c -- left).

We begin with the nonrotating configuration illustrated by the left half of the figures in Fig. 1 (with $t/U=0.06, \mu_0/U=0.7$).  There is a plateau in the superfluid density but not in the total density, and the phase of the superfluid order parameter is uniform.
 Starting from this non-rotating configuration we gradually increase the rotation speed to 
  $\nu=0.04$, iterating to self-consistency at each step. 
   Rather than forming a lattice, the resulting vortices form a ring around the central $\rho^c$ peak in Fig. 1(d).   
  This  configuration is  favored because it minimizes the sum of competing energy costs:
  the rotation favors a uniform distribution of vortices, but the single vortex energy is smallest where $\rho^c$ is low.

As seen in Fig. 1(c), the phase of the superfluid inside the ring is essentially constant.  This can be understood by an analogy with magnetostatics. The velocity $\vec{\bold v}$ obeys an analog of Ampere's law 
 $\oint \vec{\bf  v}\cdot d\vec{\bf \ell}=(h/m) N_v$, where $N_v$, the number of vortices enclosed in the contour of integration, plays the role of the enclosed current. Neglecting the discreteness of the vortices in the ring, the fluid inside is motionless, while the fluid outside moves as if all the vortices were at the geometric center of the cloud.
Even with only eight vortices our system appears to approach this limit.      
If one increases the rotation speed, one
can find a state with several concentric rings of vortices in the plateau.  
Similarly, increasing $\mu_0/U$ can can lead to multiple superfluid plateaus, each of which may contain a ring of vortices.  This structure of nested rings of vortices is reminiscent of Onsager and London's original proposal of vortex sheets in liquid helium \citep{london}.

Our second example of nontrivial vortex structures is illustrated in figure~\ref{GiantFigurePrep}, where the LDA predicts a superfluid shell surrounding a Mott core.   
Rather than forming a lattice of discrete vortices, one expects that this system will form a ``giant" vortex~\citep{PhysRevA.66.053606} when rotated: the vortices occupying the Mott region, leaving a persistent current in the superfluid shell.  The energy barriers for changing vorticity are particularly high, so we generate the rotating state in two stages.  We start with a non-rotating system ($\nu=0$) at weak coupling $\left( t/U=0.2, \mu_0/U=0.3 \right)$, gradually increasing the rotation to $\nu=0.032$, where we find the square vortex lattice illustrated in Figs. 2(b -- left) and (d).  We then adiabatically reduce  $t/U$ from $0.2$ to $0.03$.  As we reduce $t/U$, the central $\rho^c$ drops, while $\rho$ approaches unity there.  Eventually we see a Mott regime at the center of the cloud.  During the evolution, we find that 8 of the vortices escape from the edge of the trap, while four of the vortices coalesce at the center of the trap and effectively form a vortex of charge 4. Such a dense packing of vorticity would be unstable in the absence of the optical lattice.  For larger systems with higher rotation rates one finds giant vortices with larger circulation.

Due to its multiply connected topology, a ring, such as the one formed here, is one of the archetypical geometries used in theoretical discussions of superfluidity~\citep{1991LNP...394....1L}.  There are several experimental schemes for creating a ring-shaped trap~\citep{ryu:260401},
and many theoretical studies of giant vortex formation stabilized by a quadratic-plus-quartic potential~\citep{
jackson-2004-69}.  Here the multiply connected geometry is spontaneously formed by the appearance of the Mott state in the center of the cloud. As was found
by Scarola and Das Sarma~\citep{scarola:210403}, this Mott region effectively pins the vortices to the center.  

By changing $t/U$ one may study a few other interesting structures.  For example, one can engineer a situation where a central superfluid region is surrounded by a Mott ring followed by a superfluid ring.  At appropriate rotation speeds one produces a configuration which has properties of both the states seen in Fig.~\ref{RingFigurePrep} and in Fig.~\ref{GiantFigurePrep}.   One will find no vortex cores (all of the vorticity is confined to the Mott ring), the central region will be stationary, and  the outer region rotates.  

Another interesting limit is found when one decreases the  thickness of a Mott/superfluid region so much that it breaks up into a number of discrete islands.  Small Mott islands act as pinning centers, while small superfluid islands form an analog of a Josephson junction array~\citep{Cataliotti:2001wc}.  

\emph{Detection}. Vortex structures near the Mott limit may be hard to detect using {\em in-situ} absorption imaging.
As is exemplified by Fig. 1(b), the vortices do not necessarily have a great influence on the density of the cloud.  This is principally because near the Mott boundary the superfluid fraction becomes small:  even though the superfluid vanishes in the vortex core, the corresponding density may not appreciably change.
Two other pieces of physics also influence the visibility.  First, near the Mott boundary one can produce vortices with Mott cores~\citep{wu:043609}.  Depending on the bulk density, this can lead to vortices where there is no density suppression at all, or even a density enhancement.  Second, the lengthscale of the vortex core, the superfluid healing length, varies with $U/t$.  For both very large and small $U/t$ the healing length is  large, while at intermediate couplings it is comparable to the lattice spacing, possibly below optical resolution. 

We argue that the vortex structures will be much more easily imaged after time-of-flight (TOF) expansion of the cloud for time $\overline{t}$~\citep{PhysRevLett.84.806}.
The density after TOF expansion is made of two pieces -- a largely featureless incoherent background from the normal component of the gas, and a coherent contribution from the superfluid component. The coherent contribution forms a series of Bragg peaks~\citep{Greiner:2002lr}, where each peak reflects the Fourier transform of the superfluid order parameter.  Neglecting interactions during time-of-flight, and using Gaussian initial states at each lattice site, we calculate the TOF column density. In the long time limit $D_{\overline{t}}\gg R_{TF}$, where $D_{\overline{t}}=\hbar \overline{t}/m\lambda$, $R_{TF}$ is the radius of the initial cloud, and $\lambda$ is the size of each initial Wannier state, the column density of the expanding cloud is
\begin{eqnarray}
n(r,\overline{t})&=&
\rho(r,\overline{t})
\left[ 
(N-N_c) 
+ |\Lambda(r,\overline{t})|^2
\right]
\\
\rho(r,\overline{t})&=&\left(\pi D_{\overline{t}}^2 \right)^{-1} e^{-r^2/D_{\overline{t}}^2}
\\
\Lambda(r,\overline{t})&=&\sum_j \alpha_j e^{-i {\bf r \cdot r}_j/D_{\overline{t}} \lambda},
\end{eqnarray}
where 
$N$ and $N_c$ are the total number of particles and condensed particles, respectively.

\begin{figure}
\includegraphics[width=1.\columnwidth]{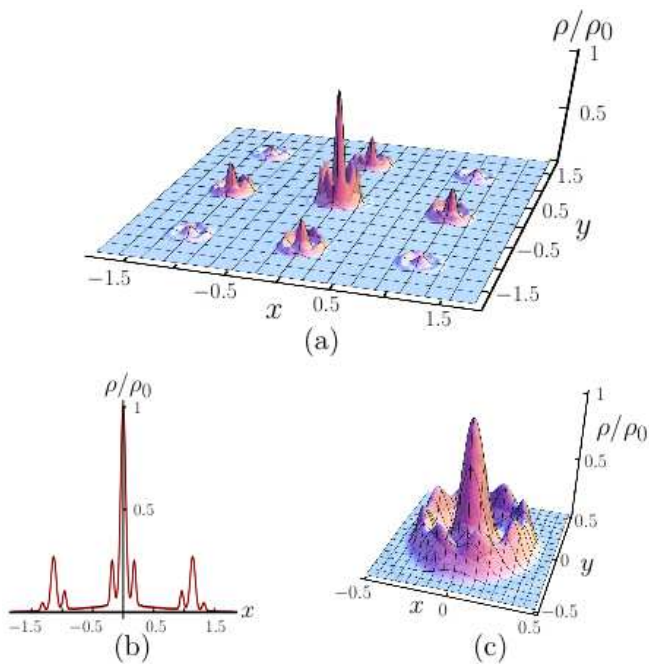}
\caption{\textbf{Time-of-flight expansion} (color online, one-column). (a) Column density $\rho$ scaled by the central column density $\rho_0$ as a function of space for the ring configuration vortex state in Fig.~\ref{RingFigurePrep} after expanding for time $\overline{t}$.  Positions are measured in terms of scaling parameter $D_{\overline{t}}=\hbar \overline{t}/m \lambda$, where $\lambda$ is the initial extent of the Wannier wavefunction.
(b) One dimensional cut through center of (a).  Note the incoherent background between the Bragg peaks. (c) Close-up of the central Bragg peak, corresponding to the Fourier transform of the superfluid order parameter.
The 8 dips in the outer crest result from  the 8 vortices in the initial state.
}
\label{TOFFig}
\end{figure}

 The incoherent contribution $(N-N_c) \rho(r,\overline{t})$ is simply a Gaussian. This is a consequence of the Gutzwiller approximation, which neglects short range correlations.
 Adding these correlations would modify the shape of the background, but it will remain smooth.  The coherent part has much more structure.
  Figure~\ref{TOFFig} illustrates the density pattern which will be seen if the rotating cloud in fig.~\ref{RingFigurePrep} is allowed to expand. 

%\section*{Summary}

%In this paper we have studied the vortex configurations in a harmonically trapped Bose gas in the presence of a rotating deep optical lattice.   We explored the influence of the Mott phase on the rotating superfluid, encountering a series of novel vortex structures such as rings and giant vortices.  Aditionally, we explained how these structures can be observed in time-of-flight images of the atomic cloud.

\emph{Acknowledgements}. We thank Joern Kupferschmidt for 
discussions about the detection method, including  some preliminary time-of-flight simulations. We thank Kaden Hazzard for providing tools for calculating the Bose-Hubbard parameters from the underlying continuum model, and for illuminating discussions.
This material is based upon work supported by the National Science Foundation through grant No. PHY-0758104.

%\bibliography{NPhys_Draft2-1}

\end{document}